\magnification = \magstep 1
\baselineskip = 14true pt

 at 12truept

\font\smrm = cmr9

\font\apj = cmcsc10
\vsize = 9.125truein
\nopagenumbers
\raggedbottom
\def\et{{\sl et al.\ }}
\def\Ha{H$\alpha $}

\def\LX{{\sl L}$_{\rm X}$}
\def\FX{{\sl F}$_{\rm X}$}

\def\E#1{$10^{#1}\, {\rm ergs\> s^{-1}}$}
\def\EE#1 #2{$#1 \times 10^{#2}\, {\rm ergs\> s^{-1}}$}
\def\FF#1 #2{$#1 \times 10^{#2}\, {\rm ergs\> cm^{-2}\> s^{-1}}$}
\def\HII {H~{\apj ii}}

\def\hi {\noindent \hangindent=2.5em}

\def\Msun{\ifmmode M_{\odot} \else $M_{\odot}$\fi}
\headline = {\tenrm\ifnum \pageno > 1 \centerline{-- \folio\ --}\else\hfil\fi}
\input epsf

\topglue 1.0truein
\centerline{THE COMPLEX BROAD-BAND X-RAY SPECTRUM}
\centerline{OF THE STARBURST GALAXY M82}

\bigskip\bigskip
\centerline{\apj Edward C.\ Moran \& Matthew
D.\ Lehnert\footnote{$^1$}{\smrm Present Address: Sterrewacht Leiden,
Postbus 9513, 2300 RA Leiden, The Netherlands}}

\bigskip\bigskip

\centerline{Institute of Geophysics and Planetary Physics}
\centerline{Lawrence Livermore National Laboratory}
\centerline{Livermore, CA 94550}

\bigskip\bigskip\bigskip
\centerline{ABSTRACT}
\bigskip

The broad-band X-ray spectrum of the prototypical starburst galaxy M82 is very
complex.  At least three spectral components are required to fit the combined
{\sl ROSAT\/} and {\sl ASCA\/} spectrum in the 0.1--10 keV range.  The
observed X-ray flux in this band is dominated by a hard $\Gamma = 1.7$,
heavily absorbed power law component which originates in the nucleus and
near-nuclear disk of the galaxy.  Among the candidates for the origin of this
hard X-ray emission, the most plausible appears to be inverse-Compton
scattered emission from the interaction of M82's copious infrared photon
flux with supernova-generated relativistic electrons.  The measured
intrinsic luminosity of the power law component agrees closely with 
calculations of the expected inverse-Compton luminosity.  Moreover,
the radio and X-ray emission in the nucleus of M82 have the same spectral
slope, which should be the case if both types of emission are nonthermal and
are associated with a common population of electrons.  The other two spectral
components, thermal plasmas with characteristic temperatures $kT \approx$
0.6 and 0.3 keV, are associated with the star formation and
starburst-driven wind in M82.  The
warmer thermal component is heavily absorbed as well and must also originate
in the central region of the galaxy.  The softer thermal component, however,
is not absorbed, and is likely to represent the X-ray emission that extends
along M82's minor axis.  The amount of absorption required in the
three-component model suggests that the intrinsic luminosity of M82 in the
0.1--10 keV band is about four times greater than its observed
luminosity of \EE {4} {40}.

\medskip\noindent
{\sl Subject headings:} galaxies: individual (M82) --- galaxies: starburst ---
X-rays: galaxies

\vskip 1.0truein

\break
\centerline{\apj 1.\ Introduction}

\bigskip
Several different sources of X-rays have been identified as contributors to
the total X-ray emission from star-forming galaxies, including stars, accreting
binary star systems, supernova remnants, diffuse hot phases of the interstellar
medium, and outflowing ``winds'' (Fabbiano 1989).  Since the radiative
processes and/or physical conditions associated with these sources differ
considerably, X-ray spectroscopy should, in principle, reveal the relative
importance of each emission component and afford us a better understanding
of the violent processes occurring within starburst galaxies.  But X-ray
observations of star-forming galaxies have not yet provided a definitive
picture of their high-energy properties.  Starburst galaxies are not
particularly luminous X-ray sources; thus, detailed information has been
available for only the closest, highest flux objects.  Moreover, spectra of
the best-studied galaxies have been difficult to interpret because
of differences and limitations in the sensitivity, spectral resolution, and
energy range of the instruments used to acquire them.  The nearby starburst
galaxy M82 (= NGC 3034), for example, has been observed with every major
X-ray mission flown over the past 15 years.  But, as summarized in Table~1,
the spectra obtained have yielded a wide variety of models for the galaxy's
integrated X-ray emission.

The {\sl ROSAT\/} and {\sl ASCA\/} observatories, with comparable sensitivity
and spectral resolution over a combined bandpass of 0.1--10 keV, offer a
possible remedy to the situation.  In this study, we present the analysis
of high signal-to-noise X-ray spectra of M82 obtained from long exposures
with both observatories.  Our objective is to ascertain a comprehensive model
for M82's broad-band X-ray spectrum and, with use of the spatial information
provided in the {\sl ROSAT\/} image, determine the physical origins of its
different components.  These results, in addition to providing new insight
into the nature of the starburst in M82, will aid the interpretation of
{\sl ROSAT\/} and {\sl ASCA\/} observations of more distant starburst galaxies,
which are likely to be far inferior to those presented herein.

\bigskip\medskip
\centerline{\apj 2.\ X-Ray Observations and Data Reduction}
\bigskip

Observations of M82 with the {\sl Einstein Observatory\/} provided clear
evidence that the X-ray emission from M82 is very extended
(Watson, Stanger, \& Griffiths 1984; Fabbiano 1988).  Unfortunately, the
imaging capabilities of the optics employed on {\sl ASCA\/} provide only
limited information about the spatially resolved emission.  Therefore, this
investigation focuses primarily on the integrated spectrum of M82, although
we will draw upon spatial information contained in the {\sl ROSAT\/} image
to interpret the results of our spectral fits.

The {\sl ROSAT\/} and {\sl ASCA\/} X-ray data for M82 were acquired from
the HEASARC archive at NASA Goddard Space Flight Center (GSFC).  M82 was
observed with the {\sl ROSAT\/} Position Sensitive Proportional Counter
(PSPC) in the 0.1--2.4 keV energy band for a total of 26.1 ksec on 28 March
and 16 October 1991.  {\sl ASCA\/} observed M82 on 19--20 April 1993 for
27.8 ksec in the 0.6--10 keV range with the two moderate-resolution Gas
Imaging Spectrometers, GIS2 and GIS3.  In the 0.4--10 keV band, exposures
totaling 16.7 ksec and 15.7 ksec were obtained with {\sl ASCA}'s
high-resolution Solid-state Imaging Spectrometers SIS0 and SIS1, respectively.
The photon event files for all five data sets were filtered using the
standard procedures for each instrument to ensure that we have included only
the cleanest data. For example, we rejected {\sl ASCA\/} data collected during
periods of enhanced background, such as those that result from passages of
the observatory through the South Atlantic Anomaly.  Data collected at times
when the geomagnetic cut-off rigidity was low (below 6 GeV $c^{-1}$) and
when the telescope optical axes were close to the Earth's limb (within
$5^{\circ}$ for the GIS or $20^{\circ}$ for the SIS) were also filtered out.
``Light curves'' for each observation were examined to certify that no
high-background data remained.

Consistent with previously published X-ray images of M82 (Watson \et 1984;
Fabbiano 1988; Bregman, Schulman, \& Tomisaka 1995), the PSPC image indicates
that M82's soft X-ray emission extends mainly along its minor axis (Fig.\ 1).
Therefore, we collected PSPC source counts within an elliptical region
oriented at PA = $140^{\circ}$ with semimajor and semiminor axes of $8'$ and
4.\negthinspace$'5$, respectively.  We estimated the background contribution
by extracting counts from $3'$-wide annular arcs (free of bright
background point sources) located outside the source region to the southwest
and northeast of the galaxy.  Subtraction of the background (10\% of the
total source-region counts) produced a PSPC spectrum with 31,180 counts.

In the GIS fields, source counts were extracted from circular regions 8$'$ in
radius centered on the nucleus of M82.  Background counts were collected in
$3'$-wide annular arcs positioned approximately the same distance off-axis
as the bulk of the emission from M82, in order to minimize the effects of
vignetting on the background measurement.  Background accounts for 7.5\% of
the counts in the source region.  The GIS2 and GIS3 spectra of M82 contain
16,499 and 16,920 background-subtracted counts, respectively.

The SIS instruments were operated in 4-CCD mode.  We collected SIS source
counts from pixels on all four chips within 8$'$ of the center of the galaxy.
We then extracted background counts from identical regions in similarly
filtered SIS ``blank sky'' images, provided by GSFC.  Background contributes
very little to the SIS spectra (just 2.2\% for SIS0 and 2.8\% for SIS1).  The
total number of background-subtracted counts is 16,222 and 12,125 for the
SIS0 and SIS1 spectra, respectively.  The substantially lower number of SIS1
counts has several causes: (1) less good exposure time was obtained with SIS1;
(2) the source in the SIS1 observation was further off-axis, and thus more
vignetted; (3) the source in the SIS1 observation was positioned closer to
the edge of the CCD, so relatively more counts were lost down the gaps between
the chips.

The PSPC, GIS, and SIS spectra were rebinned to provide a minimum of 150, 100,
and 50 counts per energy channel, respectively, ensuring that $\chi^2$ will
be a meaningful statistic for goodness-of-fit.

\bigskip\medskip
\centerline{\apj 3.\ Modeling M82's 0.1--10 Kilovolt X-Ray Spectrum}
\bigskip

Our primary objective in this study is to characterize the spectrum of
M82 over the widest possible energy range; thus, we have analyzed the
{\sl ROSAT\/} and {\sl ASCA\/} spectra simultaneously.  We have elected
to consider first the combined PSPC/GIS spectrum in order to obtain a
low-resolution overview of the broad-band X-ray properties of M82.
The energy resolution of the GIS is well-matched to that of the PSPC, so
the data over the entire 0.1--10 keV bandpass are comparably sensitive to
the details of models we apply.  These results provide a useful framework
within which to interpret the significantly more detailed SIS spectrum.

\break
\centerline{3.1.\ Low-Resolution PSPC/GIS Spectral Fits}
\medskip

Separate analyses of the PSPC and GIS spectra of M82 give very discrepant
results, primarily due to the different bandpass limits of each instrument.
A simultaneous fit to these spectra, therefore, is necessary to characterize
spectral features at both the high- and low-energy extremes of the 0.1--10 keV
range.  Using the XSPEC software, we have applied a variety of models to the
combined PSPC/GIS spectrum,
each consisting of one to three spectral components.  The results of the fits
are summarized in Table 2.  To allow for differences in the absolute
calibration of each instrument, we permitted the model normalizations for
all three data sets to be independent parameters in the fits.

Based on differences in the single-temperature thermal models derived from
the {\sl Einstein\/} IPC and MPC spectra of M82, Fabbiano (1988) suggested that
the broad-band spectrum of M82 may be intrinsically complex.  The high values
of $\chi^2$ obtained for the single-component fits to the PSPC/GIS spectrum,
thermal or nonthermal, immediately rule out such simple models and confirm
Fabbiano's speculation.  It is important to note, however, that the best-fit
parameters we find for single-component models are similar to some of the
results obtained previously with other instruments ({\sl cf}.\ Table 1).
Nonetheless, there is no question that a more complex model is required.

The presence of a very conspicuous emission line near 1.9 keV in the GIS
spectra (corresponding to Si {\apj xiii} and Si {\apj xiv}) suggests that at
least one of the components in a multiple-component fit should be an optically
thin thermal plasma (Raymond-Smith plasma, hereafter R-S).  Thus, we have tried
two-component models involving a thermal bremsstrahlung (TB) or power law
(PL) in combination with a R-S.  (Heavy element abundances in all R-S
components we employ are assumed to be solar.)  A similar two-component
model was used by Petre (1992) to fit the BBXRT spectrum of M82.  
A substantial reduction in $\chi^2$ is achieved with the addition of the
second component; the resultant ${\chi_{\nu}}^2$ ($= \chi^2/\nu$,
where $\nu$ is the number of degrees of freedom) is about 1.3.  
The ``hard'' (PL or TB) component is very hard indeed, with 
an associated photon index $\Gamma \approx 1.6$ or temperature
$kT \approx 15$ keV.  The ``warm'' R-S component has a temperature $kT
\approx 0.6$ keV.  Note, however, that the best-fit absorption column density
required for the warm component ($N_{\rm H}$ = 2.5--3.6 $\times 10^{20}$
cm$^{-2}$) is well below the value of the Galactic neutral hydrogen column
density in the direction of M82 of $N_{\rm H}$ = 4.5 $\times 10^{20}$
cm$^{-2}$ (Stark \et 1992).  Furthermore, the
fit, displayed in Figure 2$a$ with the PSPC/GIS spectra, does a
very poor job in the vicinity of the silicon lines.

To better fit the emission lines, we added a third component---a second R-S
plasma---to the model.  Once again, the overall fit improves significantly
with the additional component: $\chi^2$ decreases by more than 100.
Interestingly, the
spectral parameters for the original two components are virtually unchanged:
for the hard component, $\Gamma$ = 1.7 (or, $kT$ = 18 keV), and for the
warm component, $kT$ = 0.5--0.6 keV.  The third component is very soft, with a
temperature $kT \approx 0.3$ keV.  What has changed in the three-component
fit is the degree to which each component is absorbed.  The hard and warm
components are both absorbed by large columns ($\sim$~10$^{22}$ cm$^{-2}$).
The very soft component, on the other hand, is absorbed by a column just
equal to the Galactic value.  The PSPC/GIS spectrum with the three-component
fit is displayed in Figure 2$b$.  Comparison of Figure 2$b$ to the
two-component fit shown in Figure 2$a$ illustrates clearly the higher
quality of the three-component fit, which alone justifies adoption of
the more complex model.  We present additional evidence in \S~4.1 which
supports the three-component model in detail.

The PSPC and GIS3 spectra and the three-component model (with the instrument
responses unfolded) are displayed in Figure 3 to illustrate the relative
contribution of each component to the total X-ray emission from M82.  In this
model, the total X-ray flux in the 0.3--10 keV band is \FX = \FF {3.2} {-11},
which, at an assumed distance of 3.25 Mpc (Tammann \& Sandage 1968),
corresponds to a luminosity of \LX = \EE {4.0} {40}.  The {\sl unabsorbed\/}
flux and luminosity in the same band are four times greater than the
observed values: \FX = \FF {1.3} {-10} and \LX = \EE {1.6} {41}.  The
observed fraction of flux contained in each of the hard, warm, and soft
components is 0.68, 0.24, and 0.08, respectively.  In the absence of
absorption, these fractions would be 0.24, 0.73, and 0.03.

\bigskip
\centerline{3.2.\ The High-Resolution SIS Spectrum}
\medskip

The {\sl ASCA\/} SIS, with higher spectral resolution and greater low-energy
sensitivity than the GIS, should offer further insight into the X-ray
properties of M82.  But as Table 3 indicates, none of the model types
considered in the PSPC/GIS analysis provides a statistically acceptable fit
to the SIS spectrum.  The best fit is obtained with a three-component model
similar to the one determined in the previous section, but the fit is extremely
poor (${\chi_{\nu}}^2$ = 1.78).  However, as illustrated in Figure 4, the
discrepancy between the three-component model and data near the emission lines
in M82's spectrum dominates the contribution to $\chi^2$.  The poor SIS fit,
therefore, is {\sl not\/} an indication that the three-component model is
incorrect; instead, it reveals that R-S plasmas, which assume ionization
equilibrium, do not adequately describe M82's emission-line gas
at the resolution of the SIS.  The inclusion of additional thermal components
or nonsolar heavy element abundances does not improve the SIS fit
significantly.  Thus, nonequilibrium ionization conditions in the hot gas,
rather than multiple gas temperatures or enhanced abundances, are likely to
be responsible (Shull 1982).  Differences in the ``best fit'' spectral
parameters obtained in the SIS and PSPC/GIS fits are primarily artifacts of
the $\chi^2$ minimization fitting procedure.  For this reason, the exact
temperatures and abundances of the thermally emitting gas derived using R-S
components, which are clearly inappropriate, should be considered
uncertain (see Shull 1982).

The SIS and GIS data can be used in combination to evaluate the shape
of M82's hard X-ray spectrum, which is unaffected by absorption and
uncontaminated by the soft thermal components above $\sim$~5 keV (see Fig.\ 3),
independent of the other details of the three-component model.
A power-law fit to all four spectra in the 5--10 keV band yields a photon
index $\Gamma = 1.70 \pm 0.16$ (90\% confidence for one interesting parameter).

\bigskip\medskip
\centerline{\apj 4.\ The Origin of X-Rays in M82}

\bigskip
\centerline{4.1.\ The Nature of the Hard Spectral Component}
\medskip

The heavy absorption of the hard X-ray component found in the fit to the
PSPC/GIS spectrum of M82 implies that this component is associated with the
nuclear region of the galaxy, which was found to be more absorbed than
surrounding regions in the {\sl Einstein\/} IPC study of M82 (Fabbiano 1988).
To confirm this association, we constructed a simple X-ray ``hardness map''
from the PSPC image, which, despite its limited energy range, provides insight
into the spatial dependence of M82's spectral properties.  The hardness map
was made by dividing the hard-band (1.0--2.4 keV) PSPC image by the soft-band
(0.1--1.0 keV) PSPC image.  The demarcation energy of 1 keV was chosen because
the hard spectral component dominates the emission above $\sim$~1 keV in the
three-component model.  The hardness map, displayed in Figure 5, clearly
illustrates that the hardest X-ray flux is emitted in the nucleus and
near-nuclear disk of M82, and that the emission extended along the galaxy's
minor axis is comparatively much softer.  The hard spectral component,
therefore, must originate mainly in the nuclear region.

The hardness map also qualitatively confirms the degree to which the hard
spectral component is absorbed in the three-component model.  In the nucleus
of M82, the hardness map indicates that counts in the hard band outnumber
those in the soft band by as much as a factor of 4.  In the three-component
model, hard-to-soft counts ratios in excess of 4 are possible, even with a
considerable contribution by the $\sim$~0.5 keV thermal component.  Models
lacking significant absorption of the hard spectral component simply cannot
produce hard-to-soft counts ratios this high.  For example, in the
two-component models discussed in \S~3.1, the maximum possible ratio of
hard to soft counts is just 1.7, even if {\sl all\/} of the nuclear emission
is attributed to the hard component.

Thus, the following picture has emerged: most of the X-rays from M82 with
energies in excess of $\sim$~1 keV originate from the near-nuclear region of
the galaxy; they are associated with a hard, highly absorbed ($N_{\rm H} >
10^{22}$ cm$^{-2}$) component that is well fitted by a power law with a
photon index $\Gamma$ $\approx$ 1.7 or, equivalently, with a $\sim$ 18 keV
bremsstrahlung model.  This emission could arise from a number of possible
sources, including a buried active nucleus, an extremely hot, diffuse gas,
an ensemble of X-ray binary systems, or inverse-Compton scattered radiation.
In this section, we examine the case for each of these possibilities.

\medskip
\centerline{\sl 4.1.1.\ A Buried Active Nucleus?}
\medskip

The nuclear activity in M82, based on observations across the entire
electromagnetic spectrum, has always been attributed to a vigorous burst of
star formation.  M82's optical spectrum is unambiguously \HII\ region-like
({\sl e.g.}, Kennicutt 1992), its discrete nuclear radio sources are spatially
resolved, suggesting that they are supernova remnants (Muxlow \et 1994), and
its nuclear X-ray emission is extended (Watson \et 1984; Bregman \et 1995).
By contrast, the neighboring galaxy M81, which has an X-ray luminosity nearly
identical to that of M82, possesses all the attributes of an active
galactic nucleus (AGN): a broad emission-line optical spectrum (Peimbert \&
Torres-Peimbert 1981), a compact, inverted-spectrum nuclear radio source
(Bartel \et 1982), and a point-like nuclear X-ray source (Elvis \& Van
Speybroeck 1982).  Nonetheless, the hard $\Gamma \approx 1.7$ X-ray power-law
spectrum we find for M82 is suspiciously similar to the canonical Seyfert
galaxy X-ray spectrum ({\sl e.g.}, Nandra \& Pounds 1994).  For this reason,
we would like to consider the possibility that M82 harbors a buried active
nucleus.

Although we cannot isolate M82's nuclear X-ray source spatially in the
{\sl ASCA\/} images, the combined results of the PSPC/GIS spectral fitting
(Fig.\ 3) and the PSPC hardness map (Fig.\ 5) suggest that it can be isolated 
{\sl spectrally\/} by considering the galaxy's emission in the 3.5--10 keV
band.  A light curve of the GIS data in this energy range, displayed in
Figure~6, indicates that M82's nuclear X-ray source is not significantly
variable on timescales of minutes or hours, as some Seyfert galaxies are
known to be (see Matsuoka \et 1990).  

Measurement of M82's hard X-ray spectrum allows us to investigate whether or
not a buried active nucleus in the galaxy would be detectable at optical
wavelengths.  In Seyfert galaxies, a strong correlation between the 2--10 keV
X-ray luminosity and the broad \Ha\ emission-line luminosity has been
established (\LX /{\sl L}$_{{\rm H}\alpha} \approx$ 40; Elvis, Soltan, \& Keel
1984).  The correlation holds for the lowest luminosity Seyfert galaxies as
well (Koratkar \et 1995), including M81.  Thus, we can use the measured hard
X-ray luminosity of M82 to estimate the strength of the \Ha\ line expected
if the X-rays are produced by an active nucleus.  The unabsorbed 2--10 keV
luminosity of M82 is \EE {3} {40}, which predicts a broad \Ha\ line flux
of \FF {6} {-13}.  Assuming, however, that the emission line photons pass
through the same absorbing column as the hard X-rays, ({\sl i.e.},
$N_{\rm H} \approx 10^{22}$ cm$^{-2}$, equivalent to 5 magnitudes of
extinction at \Ha ; Zombeck 1990), the observed \Ha\ flux would be less by
a factor of 100.  In Figure 7 we have overlayed a broad \Ha\ line
of the expected intensity on an optical spectrum of M82's nucleus.  We have
fixed the velocity width of the line at 3000 km s$^{-1}$ full-width at
half-maximum, similar to \Ha\ linewidths in other low-luminosity Seyfert
galaxies (Koratkar \et 1995).  Figure 7 clearly illustrates that the broad
\Ha\ emission associated with an active nucleus would be detectable if
such a nucleus were present in M82.

Rieke \et (1980) have estimated the visual extinction toward the nucleus of M82
to be in excess of 25 magnitudes, far greater than the $\sim$~5 magnitudes
implied by the absorption of the hard X-ray component.  This discrepancy could
be due to an enhancement of the dust-to-gas ratio in M82, which, if present,
would mean that we have overestimated the broad \Ha\ flux expected from a
buried AGN.  Alternatively, the discrepancy between the visual and X-ray
extinction might suggest that the hard X-ray emission is not produced by a
compact source in the nucleus of M82, but is extended about the central
region of the galaxy.  The extended appearance of the hard X-ray source in
the PSPC hardness map (Fig.\ 5) supports this speculation.

Thus, despite the spectral similarity between M82 and a typical AGN, all
other aspects of the galaxy's nuclear emission, at any wavelength,
fail to confirm the presence of a buried Seyfert nucleus.  
We must therefore look to the starburst itself for
the source of M82's hard X-ray emission.

\medskip
\centerline{\sl 4.1.2.\ Hot, Diffuse Gas?}
\medskip

The extended X-ray halo of M82 has been interpreted as emission from hot gas
since its discovery (Watson \et 1984).  Based on the {\sl Einstein\/} IPC
spectrum, Fabbiano (1988) suggested that M82's nuclear X-ray emission
may also be thermal in nature.  As discussed in the previous section, M82's
hard X-ray source is probably distributed throughout the nuclear region.
Furthermore, we have found that a bremsstrahlung component models the
hard X-ray emission as well as a power law does, so hot gas cannot be ruled
out on spectral grounds.  However, the strong 6.7 keV Fe K$\alpha$
line {\sl expected\/} to accompany the thermal emission from a $\sim$ 18 keV
gas is not observed in the {\sl ASCA\/} spectra.  A R-S fit to the
four-instrument {\sl ASCA\/} spectrum in the 4--10 keV range indicates that
the line, if present, must be very weak, and places an upper limit to the
heavy element abundance of a gas at 0.3 solar (90\% confidence).  Hot gas
near the starburst nucleus of M82, composed of supernova ejecta and swept-up
interstellar material, is not likely to be this metal-poor (see Puxley \et
1989).

\medskip
\centerline{\sl 4.1.3.\ X-Ray Binaries?}
\medskip

By analogy to the Milky Way and other Local Group galaxies, X-ray binary
systems are expected to make an important contribution to the total X-ray
emission of star-forming galaxies (Fabbiano 1989).  {\sl ROSAT\/} HRI
observations have indicated that there are at least a few point-like X-ray
sources in M82 (Bregman \et 1995).  But if X-ray binaries are to play a
significant role in M82, they must demonstrate the appropriate spectral
characteristics.  Low-mass X-ray binaries, with fairly soft X-ray spectra
(White \et 1986)
and long evolutionary timescales ($\sim$~10$^9$ yr) relative to the age of
the starburst in M82 (10$^7$--10$^8$ yr; Rieke \et 1980; Bernl\"ohr 1993),
cannot be responsible for the majority of M82's hard X-rays.  The spectra of
high-mass X-ray binaries, on the other hand, are typically too hard ($\Gamma$ =
0.8--1.5 below 10 keV; Nagase 1989) to be compatible with M82's X-ray spectrum;
they, too, are unlikely to produce M82's hard X-ray emission.  The black hole
candidate Cyg X-1, however, possesses a low-state X-ray spectrum nearly
identical to that of M82 (Marshall \et 1993).  It would require five to ten
thousand systems emitting at Cyg X-1's low-state luminosity (several times
\E {36}) within a few hundred parsecs of the nucleus to account
for the hard X-ray luminosity of M82.  Of course, Cyg X-1's properties
may not represent the class; a typical system may in fact be more luminous,
requiring fewer such binaries to produce M82's luminosity.  On the other
hand, very few black-hole binaries have been identified {\sl anywhere\/}
in the universe; our understanding of their formation rate and lifetime
is terribly poor (Cowley 1992) and, at present, we do not know for certain
that there are {\sl any\/} black-hole binaries in M82.

\medskip
\centerline{\sl 4.1.4.\ Inverse-Compton Scattered Emission?}
\medskip

Rieke \et (1980) discussed the potential importance of inverse-Compton (IC)
scattering to the total X-ray emission from M82.  In this scenario, the
copious flux of infrared photons associated with the starburst scatters off
supernova-generated relativistic electrons.  The necessary elements for IC
scattering are certainly present in M82.  However, Watson \et (1984) dismissed
IC scattering as the dominant process based on differences between the radio
morphology, which locates the relativistic electron population, and the soft
X-ray morphology.  But as the PSPC hardness map (Fig.\ 5) indicates, the
{\sl hard\/} X-ray morphology {\sl is\/} consistent with the radio morphology.
Schaaf \et (1989) exhumed the IC hypothesis to explain the hardness of the
1.4--8.9 keV {\sl EXOSAT\/} spectrum of M82.  Although Seaquist \& Odegard
(1991) showed that the IC process contributes very little to the X-ray
emission in M82's halo, they concluded that IC losses dominate the cooling
of the relativistic electrons in the galaxy's nuclear region.

Accurate measurement of the hard X-ray spectrum of M82 permits more detailed
investigation into the role IC scattering may play.  Nonthermal radio and
X-ray emission arising from a common population of relativistic electrons
with a power law distribution of energies should have power law spectra with
the same energy index $\alpha$ (= $\Gamma - 1$).  The orientation, shape, and
size ($\sim$~$40'' \times 80''$) of the region in M82 where the 6--20 cm radio
spectral index is 0.7 or less (see Fig.\ 1 of Seaquist \& Odegard 1991) are
nearly {\sl identical\/} to that of the hard X-ray region on the PSPC hardness
map ($\sim$~$45'' \times 70''$), where the X-ray spectrum also has an energy
index of 0.7.  (The resolution of 6--20 cm spectral index map and the PSPC
hardness map are about the same.)  Furthermore, our measurement of the
intrinsic luminosity of the power law component in the 1.4--8.9 keV band of
\EE {2.5} {40} is in excellent agreement with the expected IC luminosity of
1.2--1.5 $\times$ \E {40} calculated by Schaaf \et (1989).  Taking the size
of the IC-emitting region to be 45$''$--60$''$ (suggested by the hardness map)
rather than 30$''$, as Schaaf \et assumed, brings the
agreement between the observed luminosity and the expected IC luminosity even
closer.  Since the IC emissivity depends on the product of the electron and
IR photon energy densities, the apparent nuclear confinement of M82's hard
X-rays is explained by the rapid decline of both quantities with distance
from the nucleus (Seaquist \& Odegard 1991).

We conclude that IC scattering is likely to make a significant, if not
dominant, contribution to the hard X-ray luminosity of M82.  It will require
high-resolution images at energies above a few keV, such as those the Advanced 
X-ray Astrophysics Facility ({\sl AXAF}) will provide, to determine what
fraction of the hard X-ray emission is diffuse and what fraction is produced
by discrete sources.

\bigskip
\centerline{4.2.\ The Spatially Extended X-Ray Emission}
\medskip

The three-component model of M82's X-ray spectrum indicates the presence of
hot gas in M82 emitting at two different characteristic temperatures.  The
warm 0.5--0.6 keV R-S component is heavily absorbed ($N_{\rm H} \approx
10^{22}$ cm$^{-2}$) and, like the hard X-ray source, must originate in the
central region and disk of the galaxy.  This component probably arises from
an ensemble of supernova remnants, which, individually, can have ``warm''
spectra with strong emission lines ({\sl e.g.}, Hayashi \et 1994).  At high
spectral resolution, the emission lines associated with the warm component
are poorly fitted with a simple R-S model (\S~3.2).  But R-S models often do
not accurately describe the spectra of isolated supernova remnants
({\sl e.g.}, Shull 1982; Tsunemi \et 1986), so there is no reason to expect
that a simple model should fit the integrated spectrum of a large number of
overlapping remnants.  The soft 0.3 keV R-S component, which is not
absorbed, is likely to represent the extended halo of X-ray emission
associated with the starburst-driven ``superwind'' (Bland \& Tully 1988;
Heckman, Armus, \& Miley 1990).  Leitherer, Robert, \& Drissen (1992) have
estimated that the starburst in M82 injects kinetic energy into the
interstellar medium at the rate of $\sim$~6--30 $\times$ 10$^{41}$ ergs
s$^{-1}$, well in excess of the combined intrinsic luminosity of the thermal
components in our fit (1.2 $\times$ 10$^{41}$ ergs s$^{-1}$).  For this range
of energy injection rates, the superwind model of Suchkov \et (1994)
explicitly predicts a total luminosity of $\sim$~\E {41} for the hot gas
and an extended soft X-ray component with $kT \approx$ 0.3  keV---very similar
to what we observe.  Thus, not only is it energetically feasible for the
starburst in M82 to power the observed thermal X-ray emission, the wind
generated by the starburst appears to account for the temperature and
spatial extent of the emitting gas as well.

\bigskip\medskip
\centerline{\apj 5.\ Summary}
\bigskip

We have analyzed high signal-to-noise {\sl ROSAT\/} and {\sl ASCA\/}
X-ray spectra of the starburst galaxy M82 spanning the 0.1--10 keV
energy range.  At least three spectral components contribute to the
total emission from the galaxy, revealing that its high-energy nature
is significantly more complex than previous investigations have shown.
A consistent model for the broad-band X-ray properties of M82 could not
have been determined by analyzing either data set separately.

The observed X-ray flux from M82 is dominated by a strong, hard
($\Gamma \approx 1.7$), heavily absorbed ($N_{\rm H} \approx 10^{22}$
cm$^{-2}$) power law component that originates in the nucleus and near-nuclear
disk of the galaxy.  While this spectrum resembles that of a typical
AGN, there is no other evidence to suggest that M82's hard X-rays are produced
by a buried Seyfert nucleus.  A bremsstrahlung model with $kT \approx 18$ keV
provides a good
fit to the hard X-ray component, but the strong Fe K$\alpha$ emission
expected to accompany the emission from gas at this temperature
is not observed.  The spectra of high- and low-mass X-ray binary systems
are not compatible with M82's spectrum.  The Galactic black hole candidate
Cyg X-1, however, does have a spectrum similar to that of M82, making it
plausible that back-hole binaries are responsible for M82's hard X-ray
emission. But the formation rates and lifetimes of black-hole binaries are
tremendously uncertain (Cowley 1992), and the existence of {\sl any\/} such
systems in M82 has yet to be proved.  On the other hand, the elements
required for inverse-Compton emission---infrared photons and relativistic
electrons with the appropriate energy densities---{\sl are\/} known to be
present in the nuclear region of M82.  IC emission, therefore, provides the
most straightforward explanation for the production of hard X-rays in M82.
We have shown (1) that the intrinsic luminosity of the power-law spectral
component agrees closely with the expected inverse-Compton luminosity
computed by Schaaf \et (1989), and (2) that the regions in M82 where the
radio and X-ray spectra have the same energy index are very similar
in location, size, and shape---a necessary condition if the radio and X-ray
emission are nonthermal and arise from a common population of relativistic
electrons.  We conclude, therefore, that the IC process is likely to
dominate the emission from M82's nucleus.

Also contributing to the X-ray emission from M82 are two thermal components
with characteristic temperatures of $\sim$~0.5--0.6 keV and $\sim$~0.3 keV. The
warmer of the two thermal plasmas is also heavily absorbed and may represent
supernova remnants and/or a supernova-heated phase of the interstellar medium
in the central region of the galaxy. The other thermal component is not
absorbed (above the Galactic level) and is likely to represent emission from
the starburst-driven wind that extends along M82's minor axis.  Although the
latter component contributes just a small fraction of the total X-ray flux,
its inclusion in the model is crucial for the determination of a physically
plausible picture for the X-ray emission from M82.  The thermally emitting
gas in M82 may possess a range of properties, despite the fact that it is
accurately modeled by components at two distinct temperatures at the
resolution of the {\sl ROSAT\/} PSPC and {\sl ASCA\/} GIS.  The poor fit
obtained to the high-resolution {\sl ASCA\/} SIS spectrum provides a clear
indication that the true nature of this gas is considerably more complex.
Since the spectra of individual supernova remnants are not always well-fitted
by simple R-S models (Shull 1982; Tsunemi \et 1986), it is not surprising
that a more sophisticated model is required to fit the spectrum of an
ensemble of overlapping supernova remnants.

M82 is frequently regarded as a prototype for starburst galaxies, which
gives the {\sl ROSAT\/} and {\sl ASCA\/} spectra presented here
particular importance.  Moreover, they represent the highest
signal-to-noise X-ray spectra currently available for such objects.
Their analysis, therefore, should assist the interpretation of {\sl
ROSAT\/} and {\sl ASCA\/} observations of other starbursts.
Unfortunately, until data of comparable quality are available for a
number of star-forming galaxies, the question of whether or not the
X-ray properties of M82 are prototypical remains open.

\bigskip
We are very grateful to David Helfand for a critical reading of the manuscript.
This research has made use of {\sl ROSAT\/} and {\sl ASCA\/} archival data
obtained through the High Energy Astrophysics Science Archive Research
Center (HEASARC), provided by NASA Goddard Space Flight Center.  Support for
this work was provided in part by the U.S.\ Department of Energy under
contract W-7405-ENG-48.

\break
{\topglue 1.5truein
\hsize 8truein
\epsfxsize=6.15truein
\hskip 0.15truein
\epsffile[86 281 536 590]{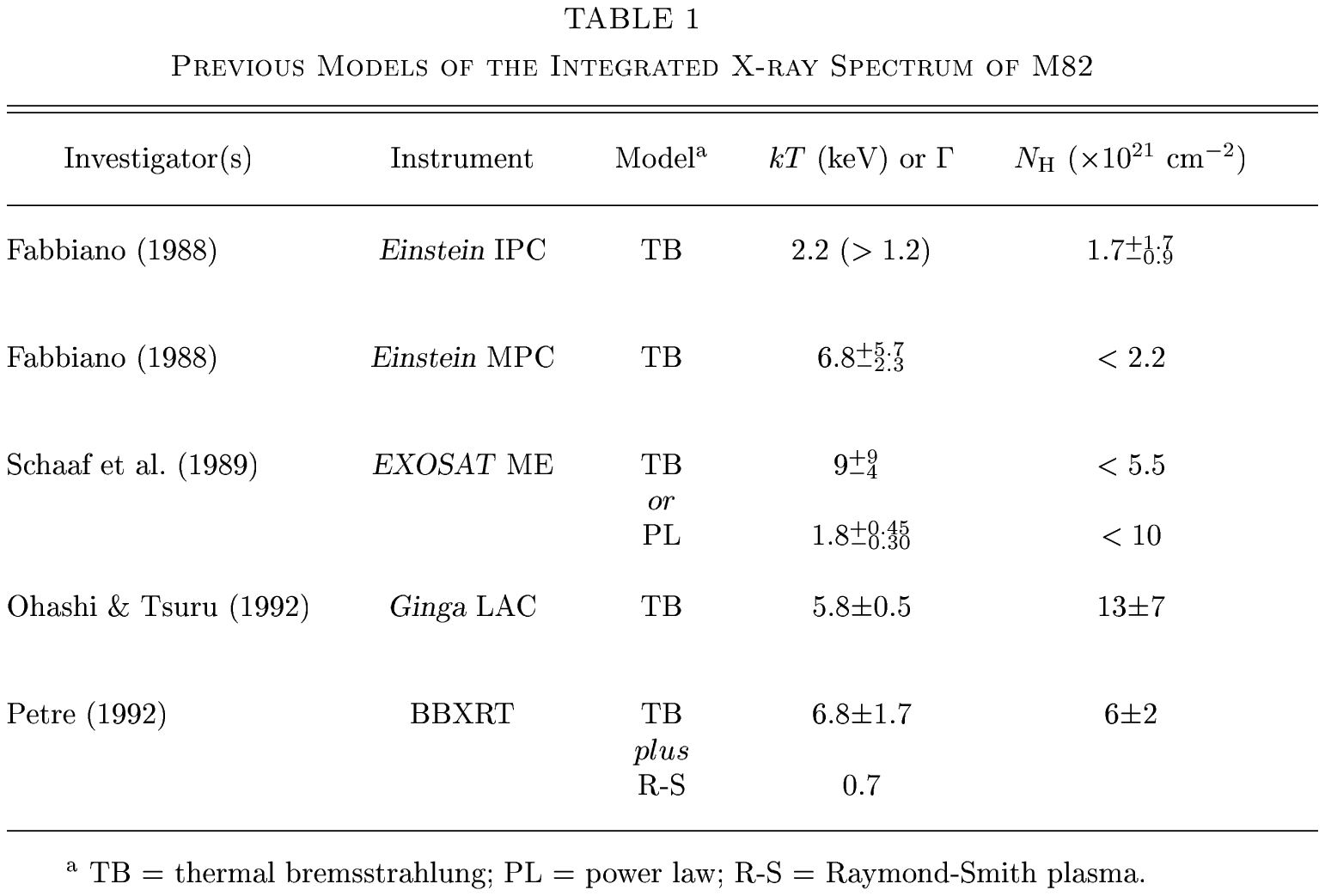}\par}

\break
{\topglue 1.5truein
\hsize 8truein
\epsfxsize=5.75truein
\hskip 0.3truein
\epsffile[100 230 515 590]{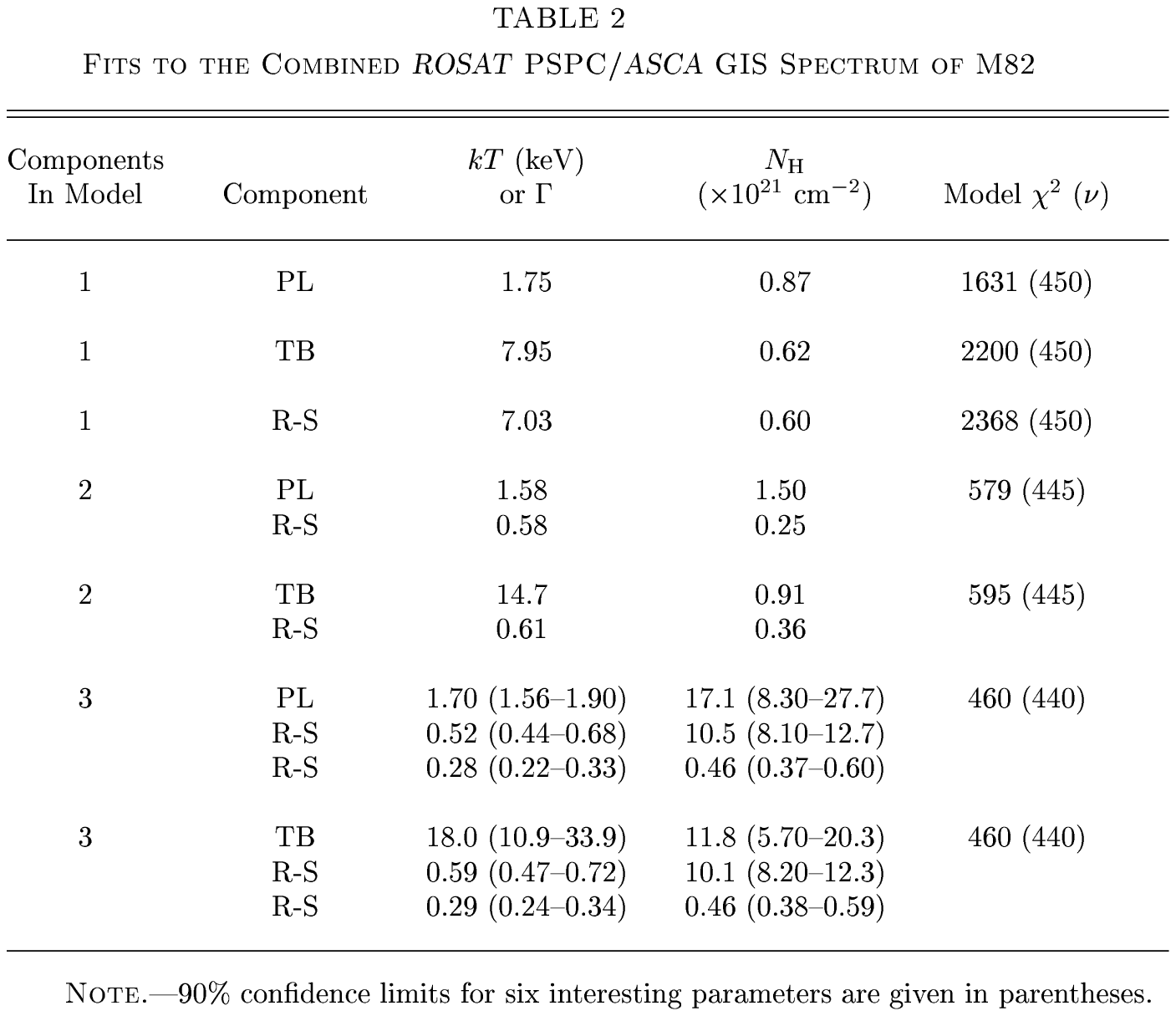}\par}

\break
{\topglue 1.5truein
\epsfxsize=4.75truein
\hskip 0.77truein
\epsffile[140 317 490 590]{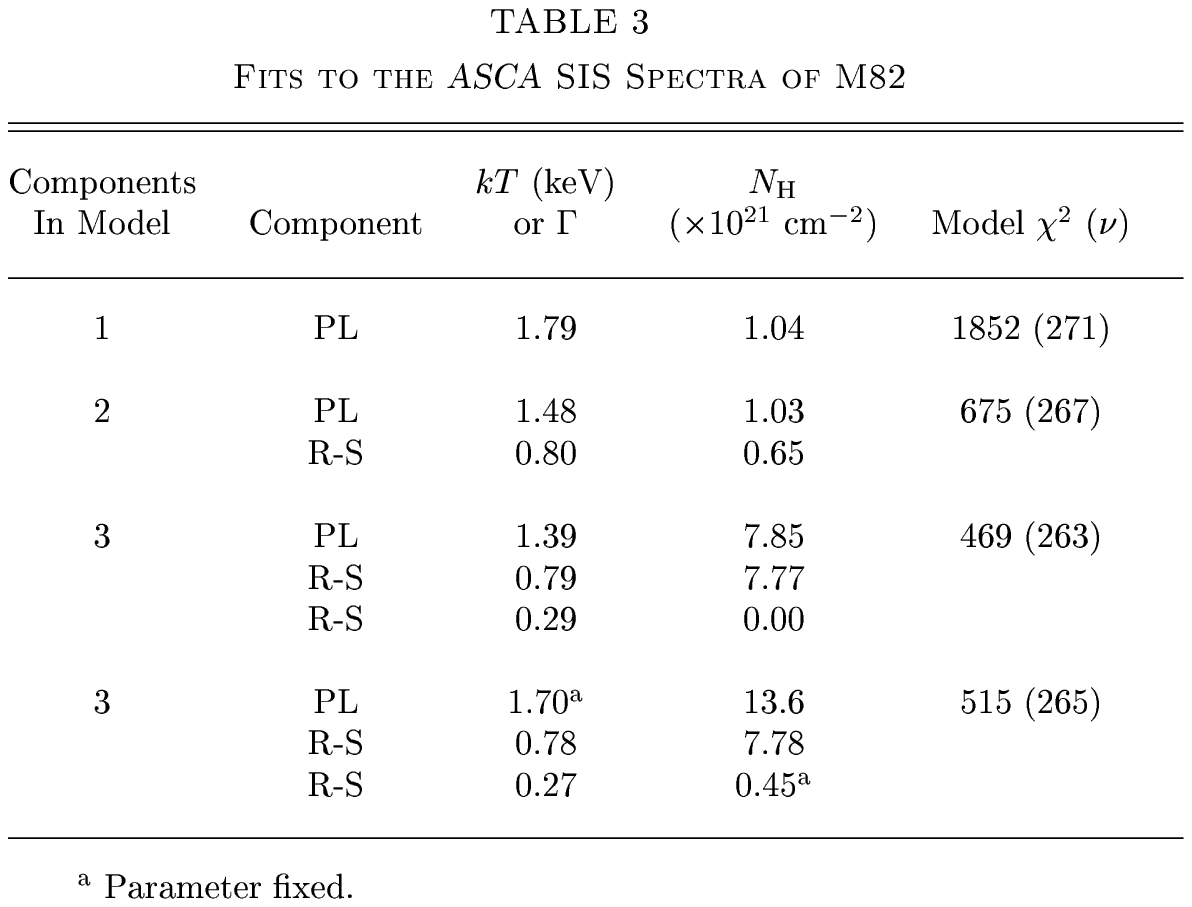}\par}

\break
\centerline{\apj REFERENCES}
\bigskip

\hi Bartel, N., \et\ 1982, ApJ, 262, 556

\hi Bernl\"ohr, K.\ 1993, A\&A, 268, 25

\hi Bland, J., \& Tully, R.\ B.\ 1988, Nature, 334, 43

\hi Bregman, J.\ N., Schulman, E., \& Tomisaka, K.\ 1995, ApJ, 439, 155

\hi Cowley, A.\ P.\ 1992, ARAA, 30, 287

\hi Elvis, M., Soltan, A., \& Keel, W.\ 1984, ApJ, 283, 479

\hi Elvis, M., \& Van Speybroeck, L.\ 1982, ApJ, 257, L51

\hi Fabbiano, G.\ 1988, ApJ, 330, 672

\hi Fabbiano, G.\ 1989, ARA\&A, 27, 87

\hi Hayashi, I., Koyama, K., Ozaki, M., Miyata, E., Tsunemi, H., Hughes,
J.\ P., \& Petre, R.\ 1994, PASJ, 46, L121

\hi Heckman, T.\ M., Armus, L., \& Miley, G.\ K.\ 1990, ApJS, 74, 833

\hi Kennicutt, R.\ C.\ 1992, ApJ, 388, 310

\hi Koratkar, A., Deusua, S.\ E., Heckman, T., Filippenko, A.\ V., Ho, L.\ C., \& Rao, M.\ 1995, ApJ, 440, 132

\hi Leitherer, C., Robert, C., \& Drissen, L.\ 1992, ApJ, 401, 596

\hi Marshall, F.\ E., Mushotzky, R.\ F., Petre, R., \& Serlemitsos, P.\ J.\ 1993, ApJ, 419, 301

\hi Matsuoka, M., Piro, L., Yamauchi, M., \& Murakami, T.\ 1990, ApJ, 361, 440

\hi Muxlow, T.\ W.\ B., Pedlar, A., Wilkinson, P.\ N., Axon, D.\ J., Sanders, E.\ M., \& de Bruyn, A.\ G.\ 1994, MNRAS, 266, 455

\hi Nagase, F.\ 1989, PASJ, 41, 1

\hi Nandra, K., \& Pounds, K.\ A.\ 1994, MNRAS, 268, 405

\hi Peimbert, M., \& Torres-Peimbert, S.\ 1981, ApJ, 245, 845

\hi Petre, R.\ 1992, in {\sl The Nearest Active Galaxies}, eds.\ J.\ Beckman,
L.\ Colina, \& H.\ Netzer (Madrid: CSIC), p.\ 117

\hi Puxley, P.\ J., Brand, W.\ J.\ L., Moore, T.\ J.\ T., Mountain, C.\ M., Nakai, M., \& Yamashita, T.\ 1989, ApJ, 345, 163.

\hi Rieke, G.\ H., Lebofsky, M.\ J., Thompson,~R.~I., Low,~F.~J., \&
Tokunaga,~A.~T.\ 1980, ApJ, 238, 24

\hi Schaaf, R., Pietsch, W., Biermann, P.\ L., Kronberg, P.\ P., \& Schmutzler,
T.\ 1989, ApJ, 336, 722

\hi Seaquist, E.\ R., \& Odegard, N.\ 1991, ApJ, 369, 320

\hi Shull, J.\ M.\ 1982, ApJ, 262, 308

\hi Stark, A.\ A., Gammie, C.\ F., Wilson, R.\ W., Bally,~J., Linke,~R.~A.,
Heiles,~C., \& Hurwitz,~M.\ 1992, ApJS, 79, 77

\hi Suchkov, A.\ A., Balsara, D.\ S., Heckman,~T.~M., \& Leitherer,~C. 1994,
ApJ, 430, 511

\hi Tammann, G.\ A., \& Sandage, A.\ R.\ 1968, ApJ, 151, 825

\hi Tsunemi, H., Yamashita, K., Masai, K., Hayakawa, S., \& Koyama, K.\ 1986, ApJ, 306, 248

\hi Watson, M.\ G., Stanger, V., \& Griffiths, R.\ E.\ 1984, ApJ, 286, 144

\hi White, N.\ E., Peacock, A., Hasinger, G., Mason, K.\ O., Manzo, G., Taylor, G.\ B., \& Branduardi-Raymont, G.\ 1986, MNRAS, 218, 129

\hi Zombeck, M.\ V.\ 1990, {\sl Handbook of Space Astronomy and Astrophysics}, (Cambridge: Cambridge University Press), pp.\ 103--104

\break
\centerline{\apj Figure Captions}

\bigskip
Fig.\ 1.---Total 0.1--2.4 keV intensity contours from the {\sl ROSAT\/} PSPC
image of M82, overlayed on an optical image of the galaxy from the POSS E
plate.  The X-ray image was smoothed using a Gaussian with $\sigma = 10''$.
Contours at the 2, 5, 10, 20, 40, 120, and 240~$\sigma$ levels are plotted.

\bigskip
Fig.\ 2.---The observed {\sl ROSAT\/} PSPC (left) and {\sl ASCA\/} GIS
(right) spectra of M82, with the best-fit ($a$) two-component and ($b$)
three-component models from Table 2 (folded through the instrument response
functions).  For these plots, the spectra have been rebinned so that the
signal-to-noise ratio in each channel is at least 12.  A power law was used
for the hard component.

\bigskip
Fig.\ 3.---The PSPC (left) and GIS3 (right) spectra of M82 and the best-fit
three-component model (with the instrument response functions unfolded)
illustrate the relative contribution of each component to the total X-ray
emission from M82.  The hard, warm, and soft components, and their sum, are
indicated with dashed, dot-dashed, dotted, and solid lines, respectively.  At
energies less than $\sim$ 0.6 keV, the soft thermal component accounts for
virtually all of the flux, whereas at energies greater than $\sim$ 1.3 keV,
the flux is dominated by the hard component.  A power law was used for the hard
component in this plot.

\bigskip
Fig.\ 4.---Detail of the SIS0 and SIS1 spectra of M82 in the 0.7--2.4 keV
range.  The spectra are fitted with the fourth model listed in Table 3.
The fit residuals indicate that the model, while fitting some of the emission
lines well ({\sl e.g.}, the Mg {\apj xii} and Si {\apj xiii} lines at 1.47
and 1.87 keV), misses badly in other cases ({\sl e.g.}, the Ne {\apj ix},
Mg {\apj xi}, and Si {\apj xiv} lines at 0.92, 1.34, and 2.01 keV).

\bigskip
Fig.\ 5.---The X-ray ``hardness map'' of M82 (grey scale), made from the ratio
of smoothed ($\sigma = 10''$) hard-band ($>$ 1 keV) and soft-band ($<$ 1 keV)
{\sl ROSAT\/} PSPC images.  The 0.1--2.4 keV intensity contours, identical to
those shown in Fig.\ 1, are overlayed.  The hard-to-soft counts ratios in the
(black) nuclear region, which measures $\sim$~$45'' \times 70''$, range from
$\sim$~2 to 4.  The halo emission extending along M82's minor axis is
comparatively much softer, with hard-to-soft counts ratios of 0.2--0.4.
The axes are labeled with J2000 coordinates.

\bigskip
Fig.\ 6.---Background-subtracted GIS2 + GIS3 light curve in the 3.5--10 keV
band, which isolates M82's nuclear X-ray emission.  In the top panel, the data
have been binned at 256 s intervals.  Data from an entire orbit have been
binned together in the lower panel.  The mean count rate is indicated with
dotted lines.  Light curves for the background are displayed at the bottom of
each panel.  No significant variations on timescales of minutes or hours are
observed.

\bigskip
Fig.\ 7.---Optical spectrum of M82's nucleus near \Ha\ $\lambda$6563, obtained
from three 30 minute observations with the Kitt Peak 4 m telescope under
photometric conditions.  A fit to the stellar continuum has been subtracted.
A broad Gaussian line with a flux of \FF {6} {-15} and a width of 3000 km
s$^{-1}$ has been overlayed at the wavelength of \Ha .  Such a line would be
expected if a Seyfert nucleus was responsible for M82's hard X-ray emission.
The vertical axis has units of ergs cm$^{-2}$ s$^{-1}$ \AA $^{-1}$.

\break
{\topglue 1.2truein
\epsfxsize=6.5truein
\epsffile{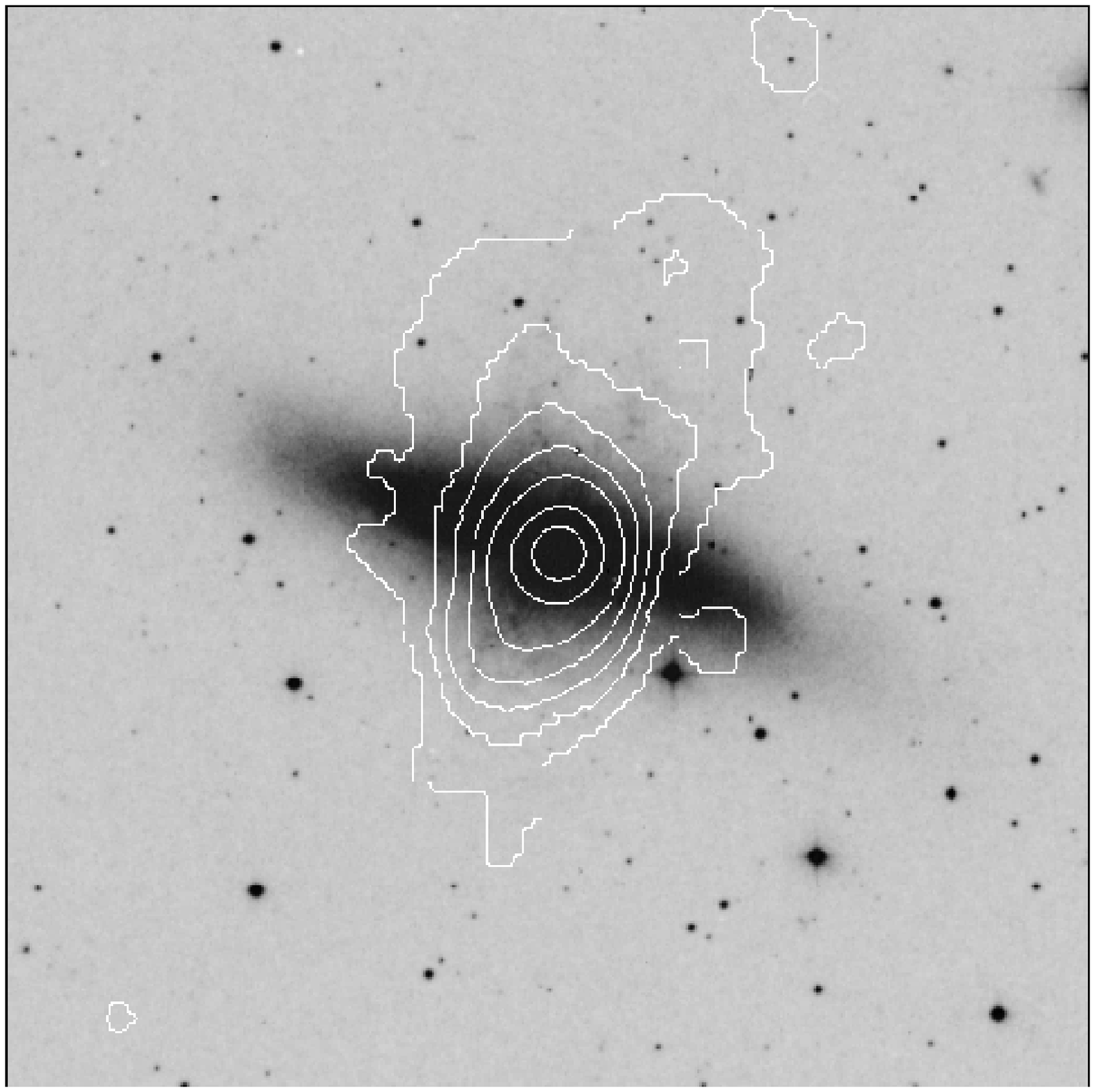}\par}
\vskip 1.0truein\centerline{Figure 1}

\break
{\topglue 0.3truein
\hsize 7.5truein
\hskip 0.1truein
\epsfxsize=5.truein
\epsffile[72 188 560 560]{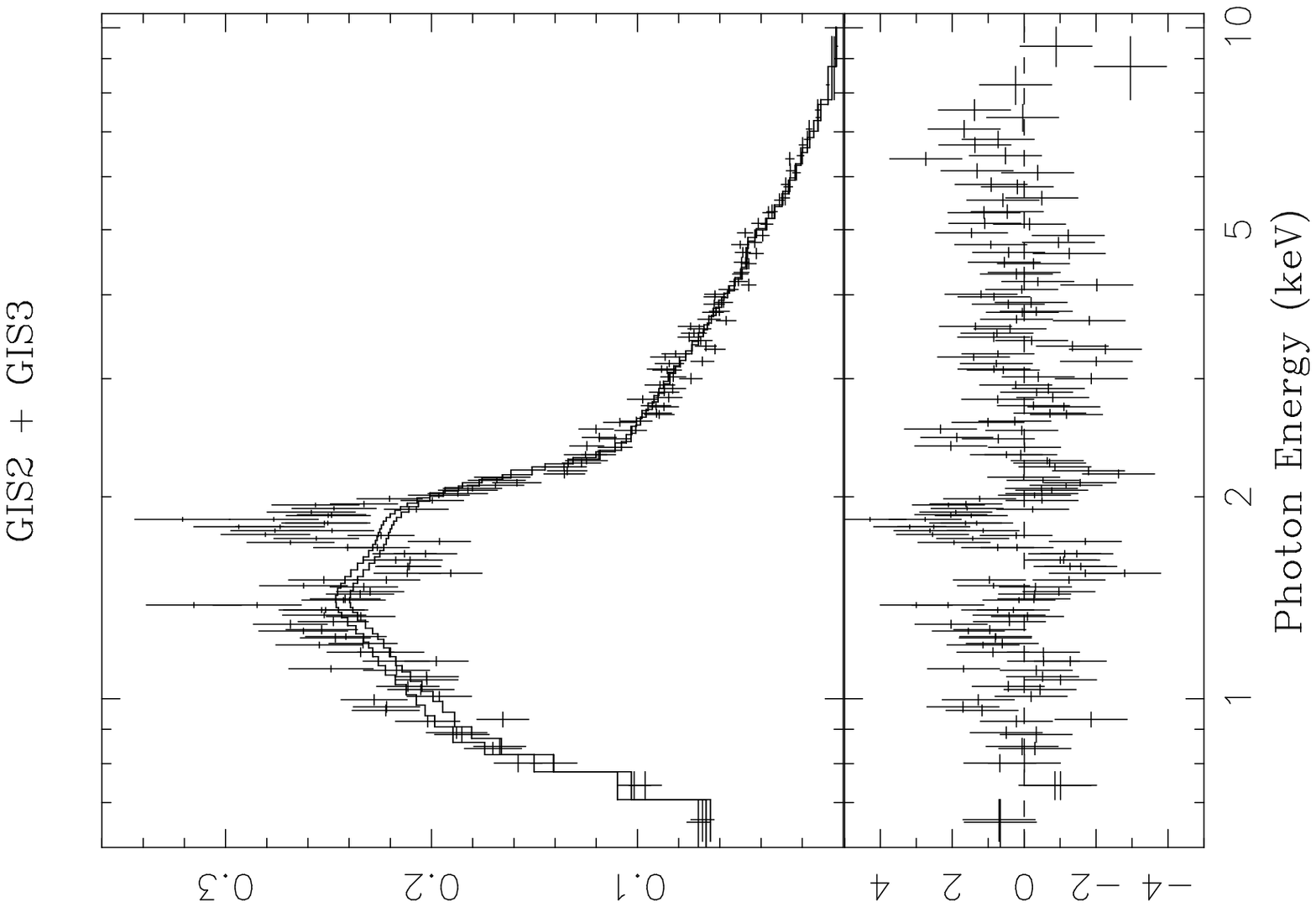}

\nobreak
\vskip -0.2truein
\hskip 0.1truein
\epsfxsize=5.truein
\epsffile[72 188 560 560]{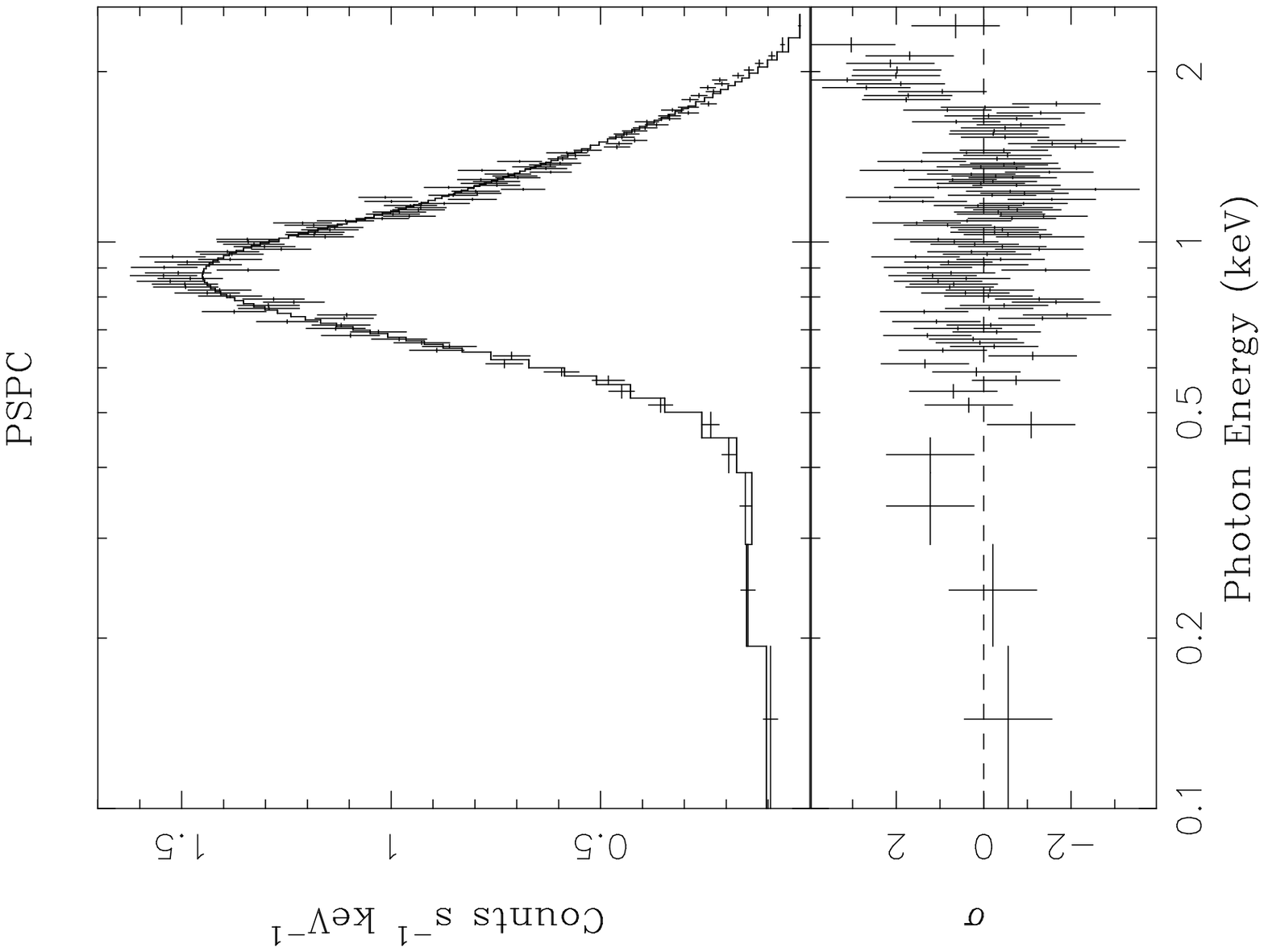}\par}
\vskip 1.0truein\centerline{Figure 2$a$}

\break
{\topglue 0.3truein
\hsize 7.5truein
\hskip 0.1truein
\epsfxsize=5.truein
\epsffile[72 188 560 560]{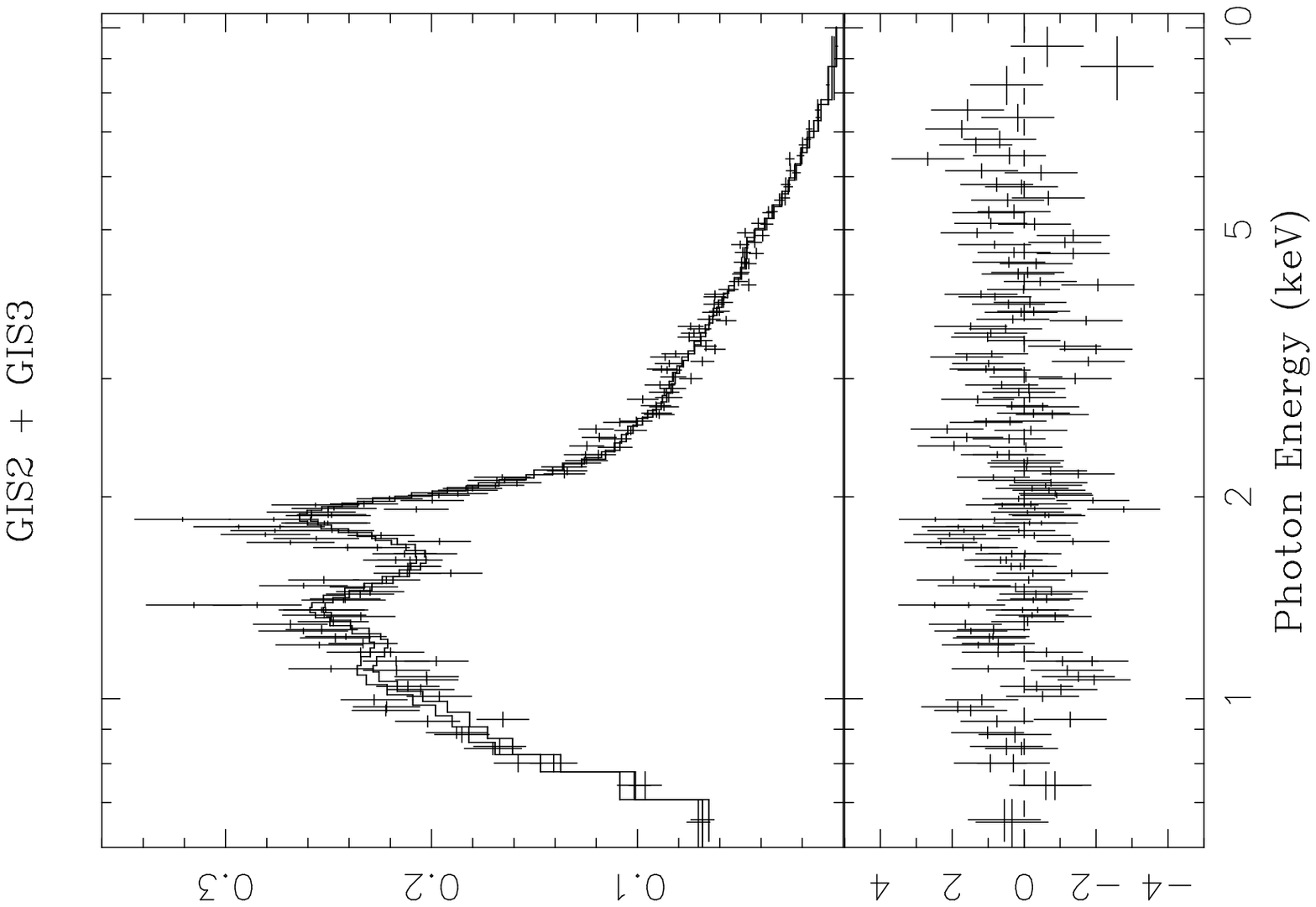}

\nobreak
\vskip -0.2truein
\hskip 0.1truein
\epsfxsize=5.truein
\epsffile[72 188 560 560]{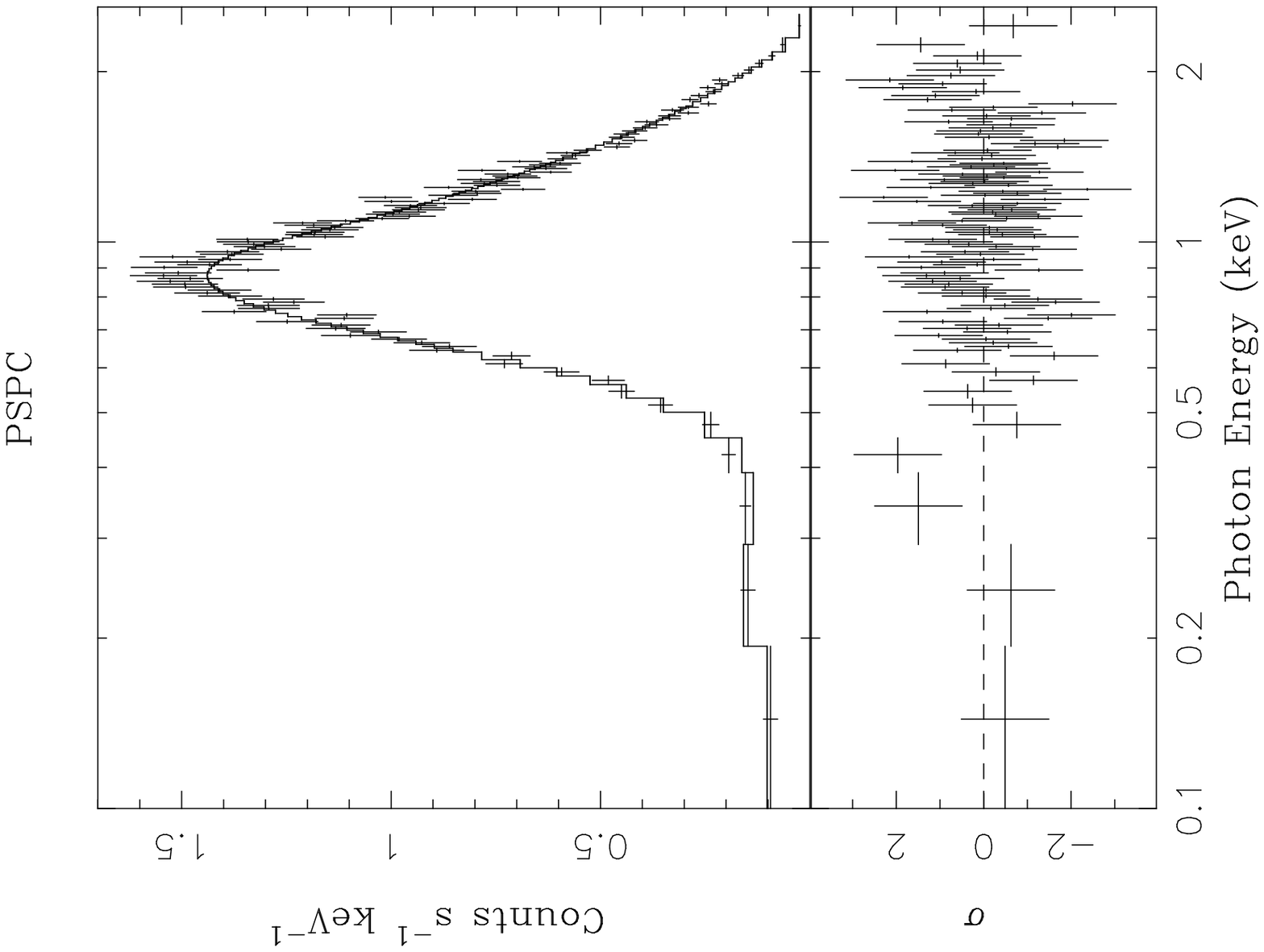}\par}
\vskip 1.0truein\centerline{Figure 2$b$}

\break
{
\hsize 7.5truein
\hskip 1.3truein
\epsfysize=4.truein
\epsffile[72 36 576 700]{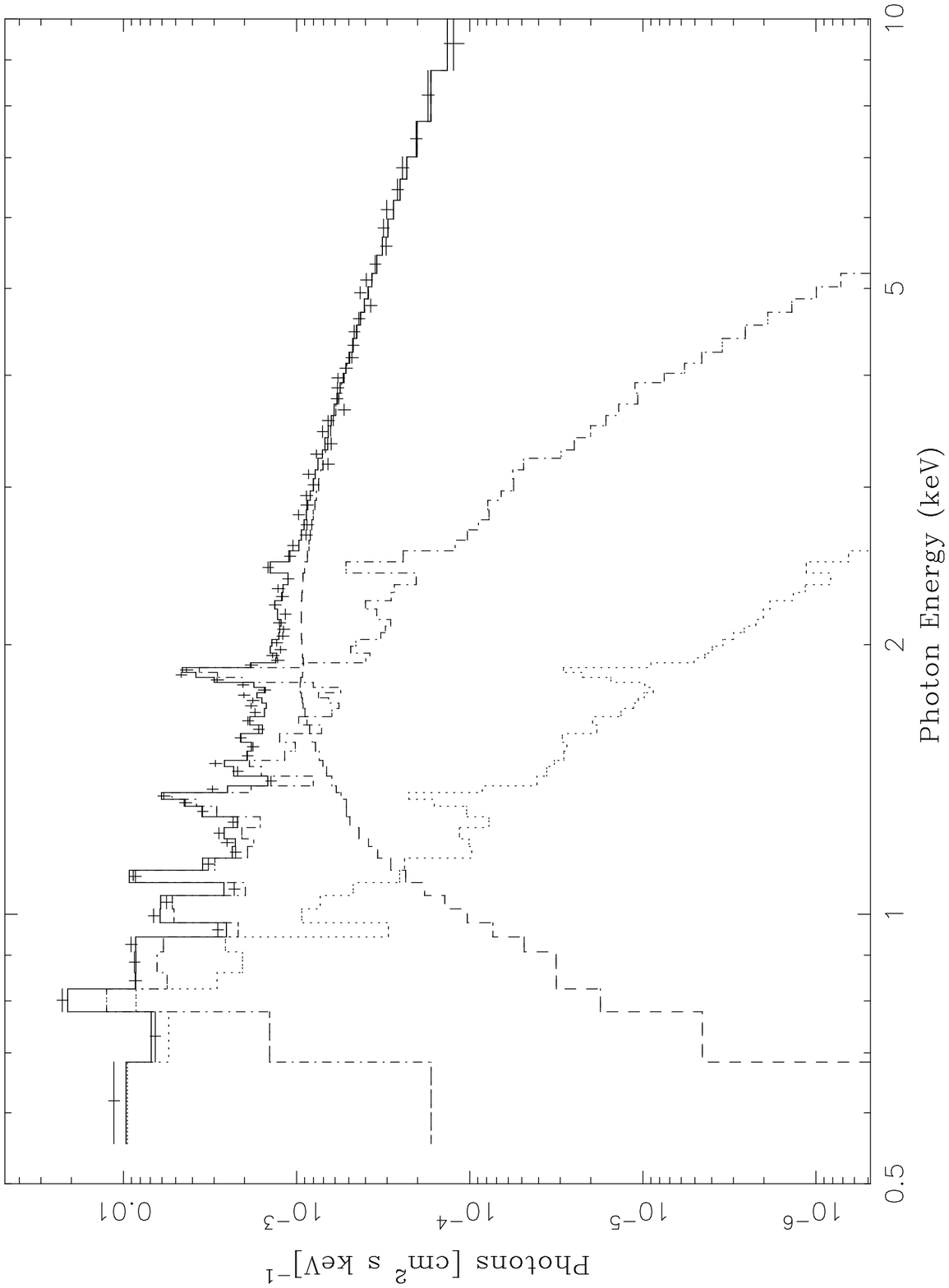}

\nobreak
\vskip 0.1truein
\hskip 1.3truein
\epsfysize=4.truein
\epsffile[72 36 576 700]{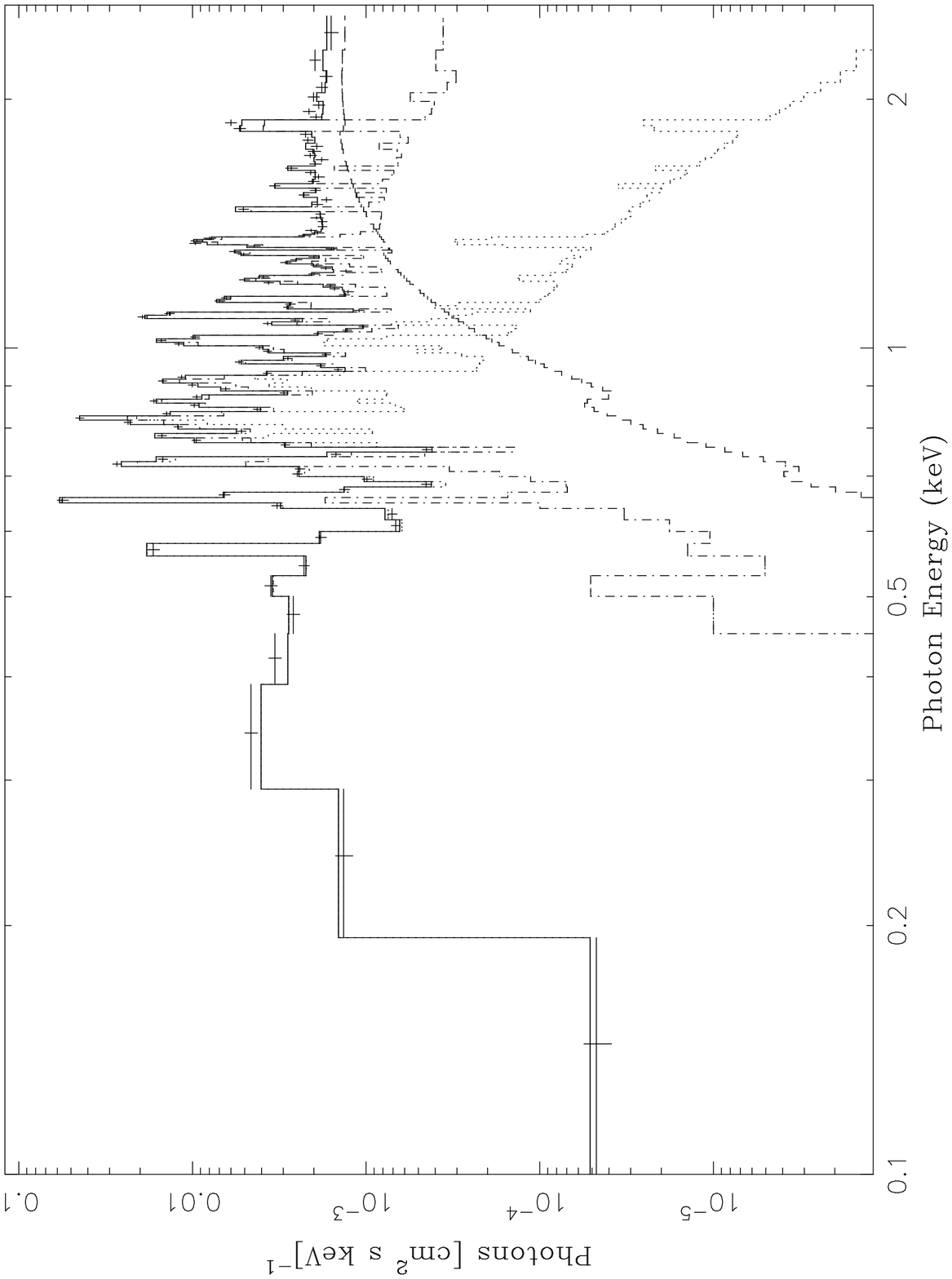}\par}
\vskip 0.8truein\centerline{Figure 3}

\break
{\topglue 1.75truein
\hsize 8.5truein
\hskip 0.2truein
\epsfxsize=6.truein
\epsffile[108 194 560 560]{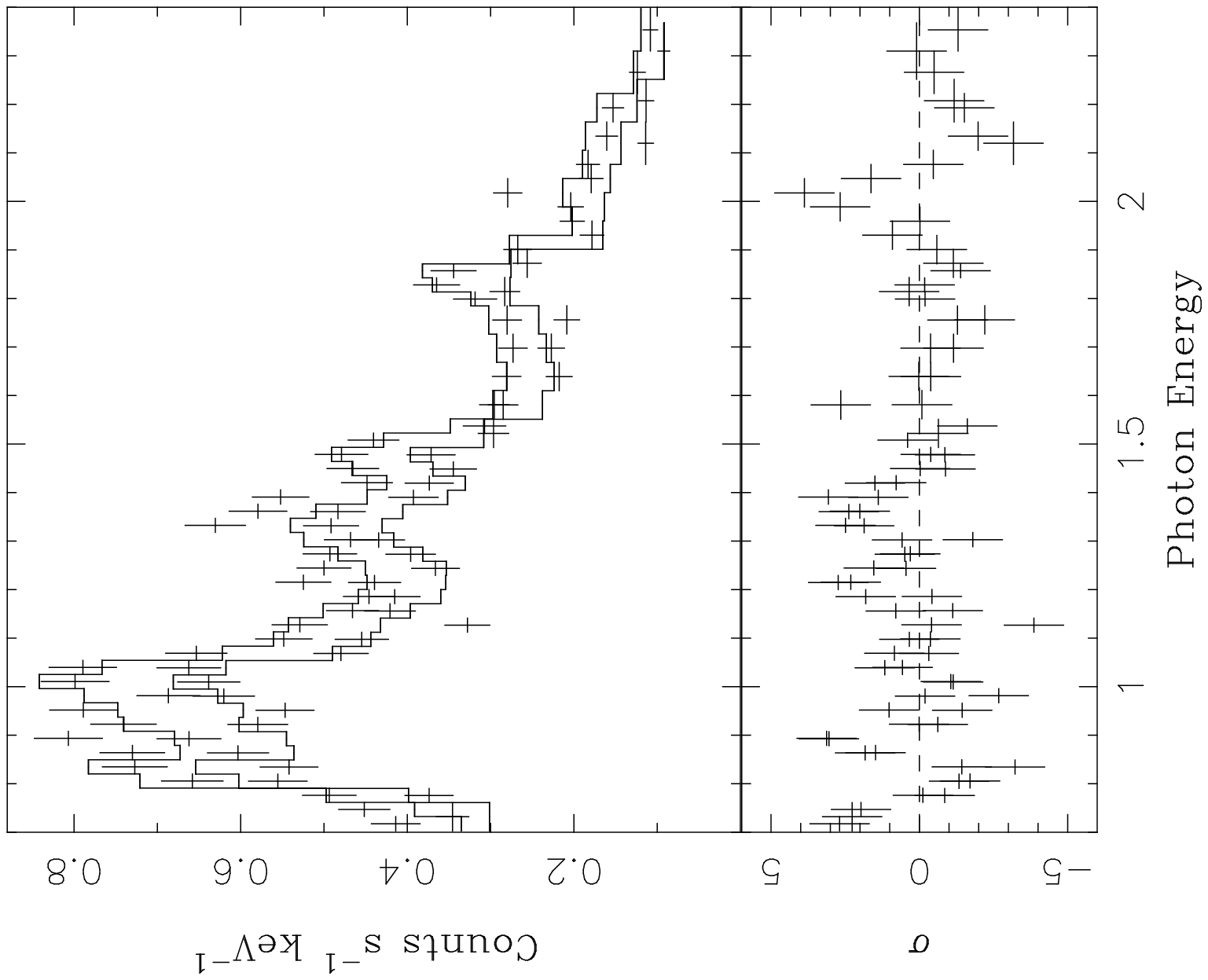}\par}
\vskip 1.0truein\centerline{Figure 4}

\break
{\topglue 1.5truein
\hskip -0.5truein
\epsfxsize=6.5truein
\epsffile{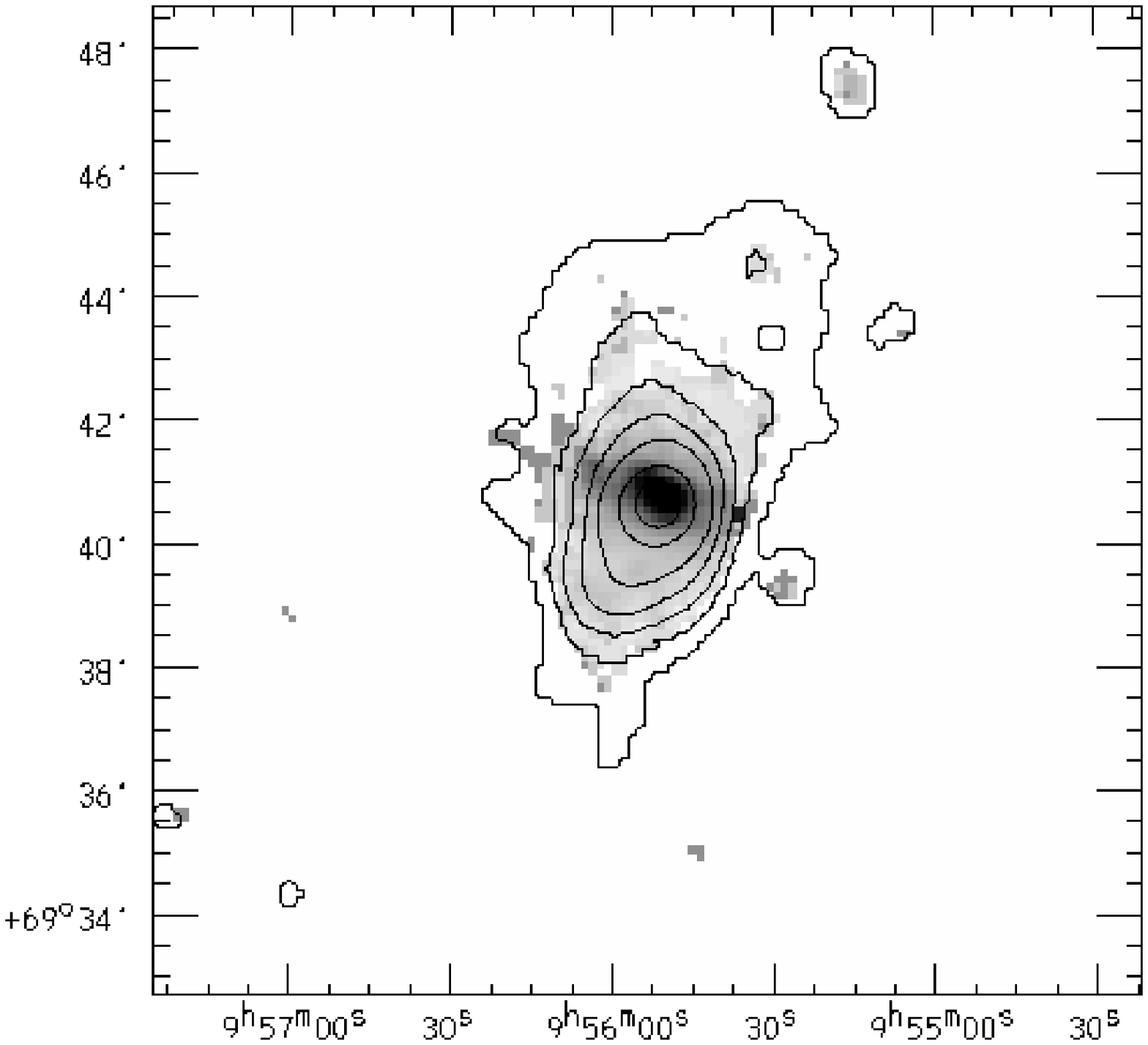}\par}
\vskip 1.0truein\centerline{Figure 5}

\break
{\topglue 0.7truein
\hskip 0.2truein
\epsfxsize=5truein
\epsffile[21 345 540 720]{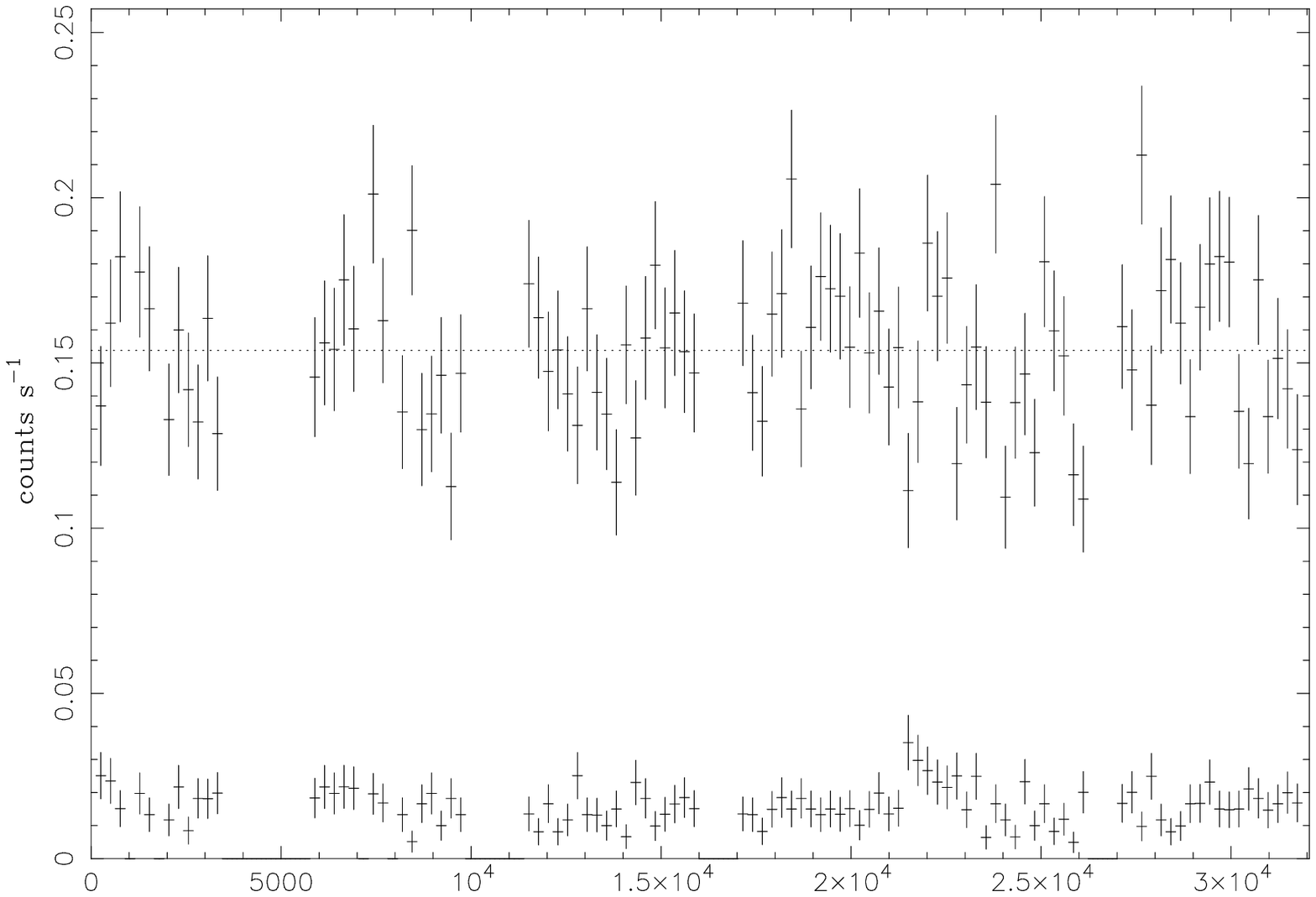}

\nobreak
\vskip -0.1truein
\hskip 0.2truein
\epsfxsize=5truein
\epsffile[21 345 540 720]{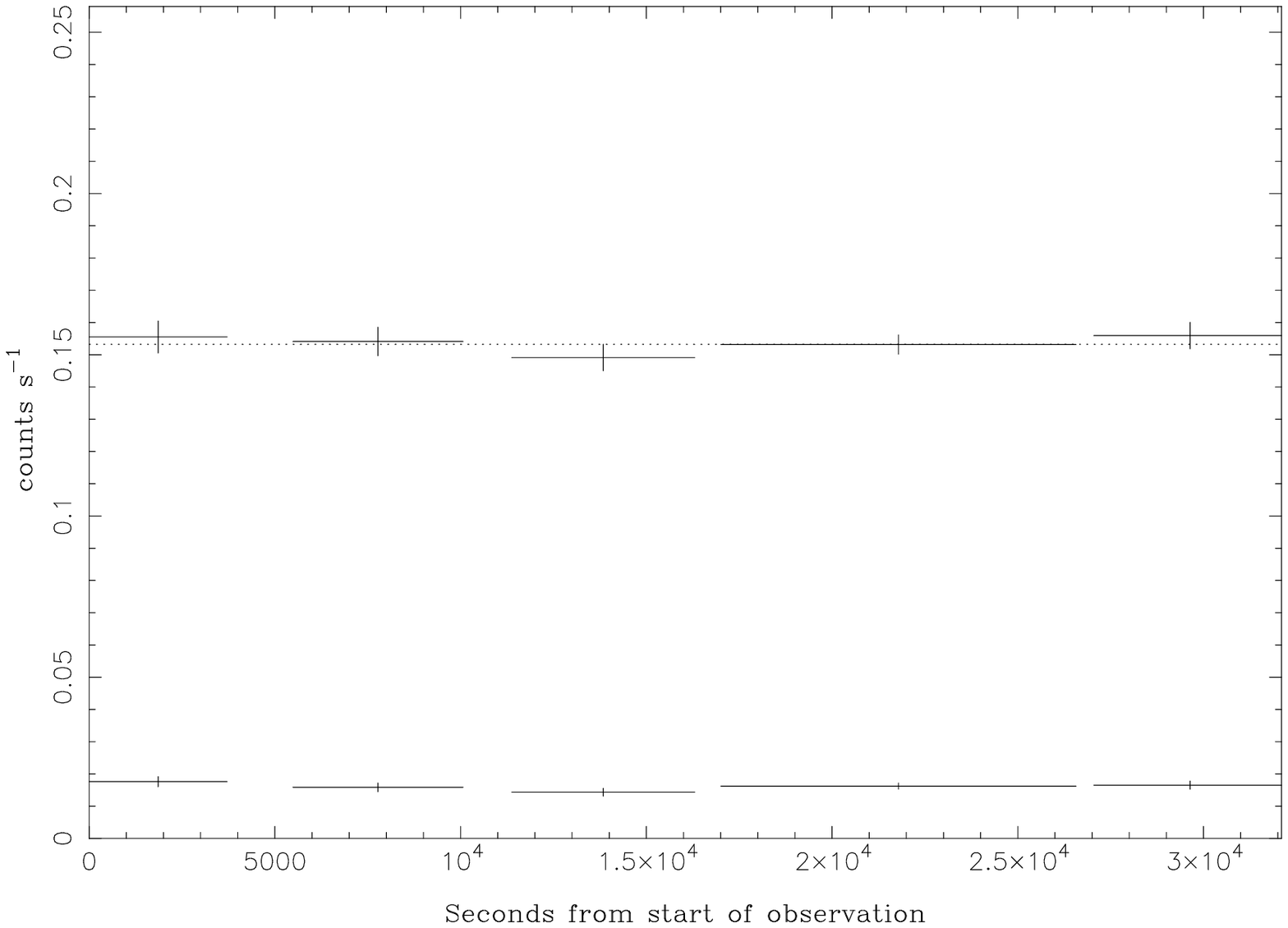}\par}
\vskip 1.0truein\centerline{Figure 6}

\break
{\topglue 2.25truein
\hskip -0.9truein
\epsfxsize=6.5truein
\epsffile[36 252 492 528]{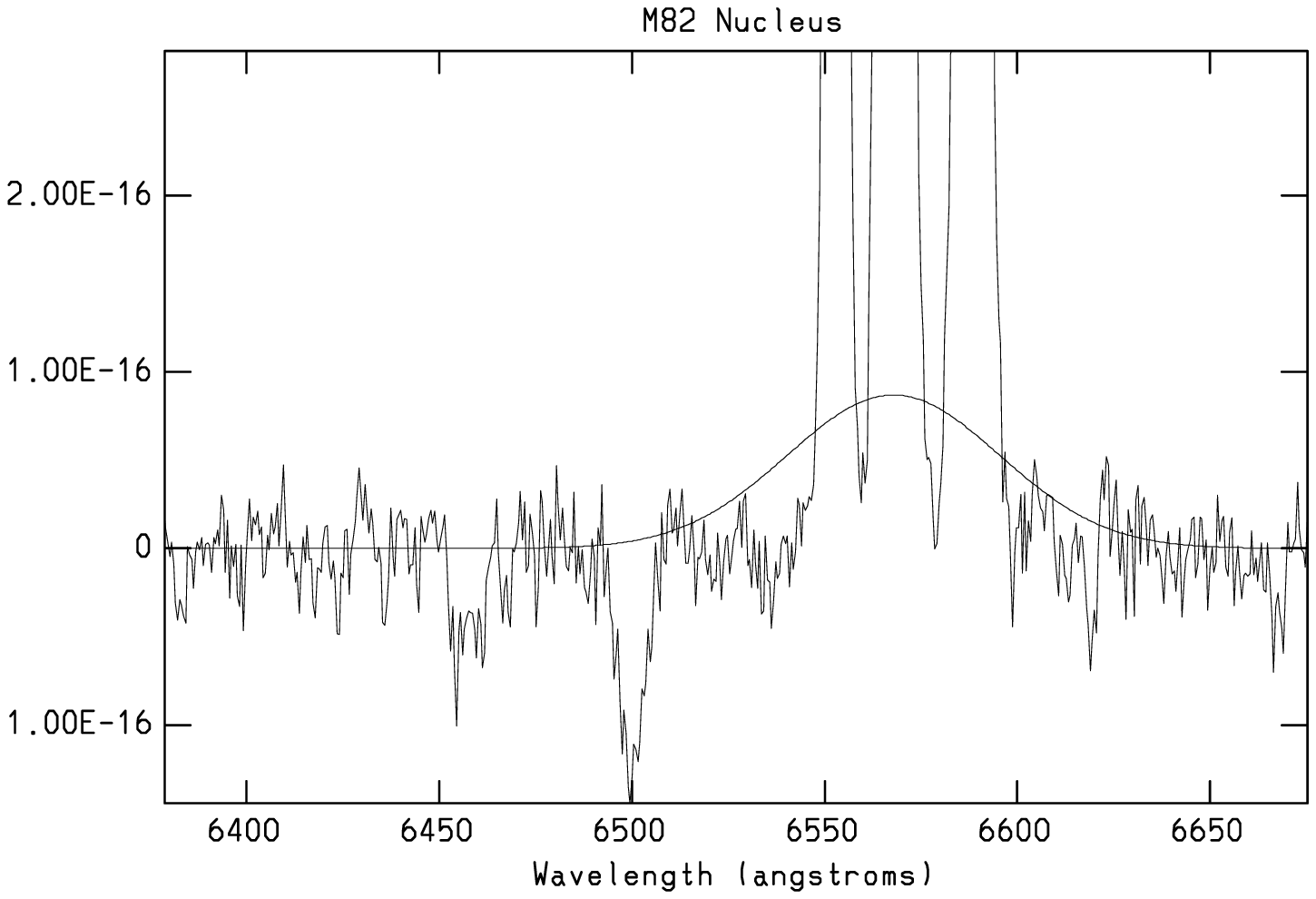}\par}
\vskip 1.0truein\centerline{Figure 7}

\bye